# Design of controllable magnon frequency comb in synthetic ferrimagnets

Y. Liu[1*], T. T. Liu[1*], Q. Q. Yang[1], G. Tian[1], Z. P. Hou[1], D. Y. Chen[1], Z. Fan[1], M. Zeng[1], X. B. Lu[1], X. S. Gao[1], M. H. Qin[1,†], and J. –M. Liu[1,2]

[1]*Guangdong Provincial Key Laboratory of Quantum Engineering and Quantum Materials and Institute for Advanced Materials, South China Academy of Advanced Optoelectronics, South China Normal University, Guangzhou 510006, China*

[2]*Laboratory of Solid State Microstructures, Nanjing University, Nanjing 210093, China*

[*]These authors contributed equally to this work
[†]Email: qinmh@scnu.edu.cn

**[Abstract]** Magnon frequency comb provides opportunities for exploring magnon nonlinear effects and measuring the transmission magnon frequency in magnets, whose controllability becomes vital for modulating the operating frequency and improving the measurement accuracy. Nevertheless, such controllable frequency comb remains to be explored. In this work, we investigate theoretically and numerically the skyrmion-induced magnon frequency comb effect generated by interaction between the magnon excitation mode and skyrmion breathing mode in synthetic ferrimagnets. It is revealed that both the skyrmion breathing mode and the magnon frequency gap closely depend on the net angular momentum $δ_s$, emphasizing the pivotal role of $δ_s$ as an effective control parameter in governing the comb teeth. With the increase of $δ_s$, the skyrmion size decreases, which results in the enlargement of the breathing frequency and the distance between the comb teeth. Moreover, the dependences of the magnon frequency gap on $δ_s$ and the inter-layer coupling allow one to modulate the comb lowest coherent frequency via structural control. Consequently, the coherent modes generated by the comb may range from gigahertz to terahertz frequencies, serving as a bridge between microwave and terahertz waves. Thus, this work represents a substantial advance in understanding the magnon frequency comb effect in ferrimagnets.

# I. INTRODUCTION

The development of the optical frequency comb provides a precise and direct link between microwave and optical frequencies, which represents an established milestone for visible/near-infrared metrology [1]. This great breakthrough has inspired scientists to search for frequency combs in other physical systems for potential applications. For example, as the quantum of collective spin excitations in ordered magnets, magnons are drawing more and more attention as information carriers for their low power consuming and convenient manipulation [2–8]. Actually, many photon-like phenomena including Kerr nonlinearities and Goos-Hänchen effect have been reported in magnonic systems [9–11], as well as the magnon frequency comb effect.

Specifically, the magnon frequency combs have been predicted in earlier theoretical works, which primarily relies on the nonlinear processes of magnons attributing to the three-magnon effect in magnetic media [12–14]. However, the three-magnon interaction of uniform states are typically weak, which poses a significant challenge in creating magnon frequency combs with sufficiently good performance. Interestingly, such nonlinear interaction can be significantly enhanced via the scattering between magnons and noncollinear magnetic textures or so-called magnetic solitons [12,15–17], which certainly benefits to the magnon frequency combs.

Of particular interest here, are magnetic skyrmions. As a type of vortex-like magnetic textures, skyrmions have shown great promise for both fundamental physics and potential applications in spintronics [18–26]. Importantly, the nonlinear processes between the skyrmion breathing modes and spin-wave modes can generate the magnon frequency combs in ferromagnets [12]. In detail, magnons are scattered by the skyrmion and they simultaneously interact with the skyrmion breathing mode. This dual interaction gives rise to a sequence of magnon confluence and splitting processes, which ultimately results in the magnon frequency comb effect with the teeth distance corresponding to the skyrmion breathing mode. Furthermore, the skyrmion-induced frequency combs in antiferromagnets can extend the frequency range to terahertz [16,17], significantly advancing terahertz technology applications in ultrafast magnonics.

These important reports present an opportunity to realize frequency combs in magnonic devices, which extensively expands the current understanding of fundamental physics.

However, no reliable solution for regulating the comb teeth is available due to the difficulty in controlling the skyrmion breathing mode in ferromagnetic and antiferromagnetic media, as far as we know. Undoubtedly, modulating the comb teeth is essential for future applications because it determines the measurement accuracy. Furthermore, the magnon frequency combs in ferromagnets and antiferromagnets operate in the microwave and terahertz bands, respectively, while the frequency gap between the two bands definitely deserves to be bridged to broaden the application range of the comb.

In this regard, ferrimagnets which combine the benefits of ultra-high working frequencies with ease of detecting and modulating are greatly potential in solving this problem. As one of the most important parameters in ferrimagnets, the net angular momentum $\delta_s$ significantly affects the dynamics of magnons and skyrmion, and it can be elaborately adjusted through tuning temperature or material composition [27–33]. Generally, variations in $\delta_s$ are accompanied by alternations of the total magnetization, which affects the skyrmion size and in turn changes the breathing mode of the skyrmion. In addition, the critical frequency beyond which magnon excitation is possible also depends on $\delta_s$, allowing one to modulate the lowest frequency within the comb. More importantly, magnon excitation frequency in ferrimagnets reaches into hundreds of gigahertz range, allowing the comb coherence mode to be extended up to terahertz and down to microwave frequencies, as depicted in Fig. 1(a). This merit will establish a pivotal connection between the microwave and terahertz frequency bands. Thus, effective control of magnon frequency combs could be realized in ferrimagnets which urgently deserves to be clarified considering its importance for the developments of magnonics and spintronics.

In this work, we investigate numerically and theoretically the magnon frequency comb effect in ferrimagnets induced by the interaction between the magnon excitation mode and the skyrmion breathing mode. It will be demonstrated that both the skyrmion breathing mode and magnon frequency gap depends on the net spin angular momentum $\delta_s$, allowing one to control the comb through tuning $\delta_s$. In addition, the coupling between the two sublattices also affects the magnon frequency gap and modulates the lowest coherent mode of the comb. The coherent modes generated by the comb exist from gigahertz to terahertz frequencies, serving as a bridge between microwave and terahertz waves.

## II. THEORETICAL ANALYSIS

*A. Ferrimagnetic skyrmion size*

As revealed in the earlier works, skyrmion size mainly determines the energy to induce fluctuation and affects the breathing mode of skyrmion [34–36]. Considering its important role in skyrmion dynamics, it would be of first priority to theoretically derive the skyrmion size. Here, the size is estimated based on the 360° domain wall profile which has been successfully used in ferromagnets and antiferromagnets [37,38].

Without loss of generality, we study a synthetic ferrimagnetic bilayer film consisting of two diverse ferromagnetic layers with an antiferromagnetic inter-layer Ruderman-Kittel-Kasuya-Yosida exchange coupling related to the spacer, as depicted in Fig. 1(b) [39]. The continuum model Hamiltonian density of the system can be written as

$$H = A(\nabla \mathbf{m}_i)^2 + D\left[m_i^z \nabla \cdot \mathbf{m}_i - (\mathbf{m}_i \cdot \nabla)m_i^z\right] + K\left[1 - (\mathbf{m}_i \cdot \hat{\mathbf{e}}_z)^2\right] + H_{\text{demag}} + H_{\text{inter}}, \quad (1)$$

where $\mathbf{m}_i$ represents the local unit magnetization vector with the layer index $i$. $A$, $D$, and $K$ are the intra-layer ferromagnetic exchange, Dzyaloshinskii-Moriya (DM) interaction, and perpendicular magnetic anisotropy constants, respectively. $H_{\text{demag}}$ is the demagnetization energy, and the last term is the inter-layer exchange coupling between the nearest neighbors which reads

$$H_{\text{inter}} = \frac{\sigma}{c}(1 - \mathbf{m}_i \cdot \mathbf{m}_j), \quad (2)$$

where $\sigma$ is the bilinear surface exchange coefficient between the two surfaces, $c$ is the thickness of ferrimagnetic layer. Here, the dipolar interaction is ignored considering that the net magnetization of ferrimagnet could be typically several orders of magnitude smaller than ferromagnet [29,40].

Next, we introduce the Néel vector $\mathbf{n} = (\mathbf{m}_1 - \mathbf{m}_2)/2$ and the magnetization vector $\mathbf{m} = (\mathbf{m}_1 + \mathbf{m}_2)/2$ to deal with the dynamic equations of ferrimagnets. Therefore, the Hamiltonian density can be expressed as

$$H = 2A(\nabla \mathbf{n})^2 + 2D\left[\nabla n_z \cdot \mathbf{n} - (\mathbf{n} \cdot \nabla)n_z\right] + 2K(1 - n_z^2) + H_{\text{demag}} + \frac{2\sigma}{c}(1 - \mathbf{m}^2). \quad (3)$$

It is convenient to use the polar coordinates $(r, \phi)$, the polar angle $\Theta = \Theta(r, \phi)$, and the azimuthal angle $\Phi = \Phi(r, \phi)$, considering the rotational symmetry of skyrmion. A skyrmion centered at $r = 0$ can be described by

$$\Theta = \Theta(r), \quad \Phi = v\phi + \lambda, \tag{4}$$

where $v$ is the skyrmion vorticity, and $\lambda$ is the skyrmion helicity. Then, the free energy of the system including a skyrmion in the polar coordinates can be written as

$$E = 4\pi Ac \int \left[ \left(\frac{d\Theta}{dr}\right)^2 + \left(\frac{v}{r}\sin\Theta\right)^2 \right] rdr + 4\pi K'c \int \sin^2\Theta\, rdr - 4\pi\sigma \int rdr$$
$$- 4\pi Dc \int \left[ \frac{d\Theta}{dr}\cos(\Phi-\phi) + \frac{v}{2r}\sin 2\Theta \cos(\Phi-\phi) \right] rdr. \tag{5}$$

Here, $K' = K - \mu_0 M_d/2$ is the corrective easy-axis anisotropy with $M_d = (M_1^2 + M_2^2)/2$ and the saturation magnetization $M_1$ (GdCo) and $M_2$ (Co), in which the shape anisotropy of the demagnetizing field is taken into account. For a Néel type skyrmion ($v = 1$ and $\lambda = 0$), Eq. (5) is simplified to

$$E = 4\pi Ac \int \left[ \left(\frac{d\Theta}{dr}\right)^2 + \left(\frac{1}{r}\sin\Theta\right)^2 \right] rdr + 4\pi K'c \int \sin^2\Theta\, rdr$$
$$- 4\pi Dc \int \left[ \frac{d\Theta}{dr} + \frac{1}{2r}\sin 2\Theta \right] rdr. \tag{6}$$

It is noted that the inter-layer nearest neighboring spins keep antiparallel with each other when a strong inter-layer coupling is considered. Then, based on the 360° domain wall profile [37,41], the skyrmion size can be analytically calculated by

$$R = \pi D \sqrt{\frac{A}{16AK'^2 - \pi^2 D^2 K'}}. \tag{7}$$

It is shown that the skyrmion size closely depends on the corrective anisotropy $K'$, in addition to the exchange and DM interactions. As a result, one may tune the net angular momentum $\delta_s$ and the total saturation magnetization to control the skyrmion size, which in turn modulates the breathing modes of the skyrmion [34,42].

*B. The three-magnon processes*

The interaction between the breathing mode of skyrmion and magnons should be clarified to understand the three-magnon process which is responsible for the frequency comb effect. We introduce bosonic creation ($a_k^\dagger$) and annihilation ($a_k$) operators and reformulate the Néel vector **n** using the local coordinate ($\hat{e}_1$, $\hat{e}_2$, $\hat{e}_3$) and the Holstein-Primakoff transformation [12,43],

$$n_1 \approx \frac{1}{\sqrt{2S}}\left[a+a^\dagger - \frac{a^\dagger aa + a^\dagger a^\dagger a}{4S}\right],$$
$$n_2 \approx -\frac{i}{\sqrt{2S}}\left[a-a^\dagger - \frac{a^\dagger aa - a^\dagger a^\dagger a}{4S}\right], \quad (8)$$
$$n_3 \approx 1 - a^\dagger a.$$

By substituting these connections back into Eq. (1), we derive the three-magnon terms as

$$H^{(3)} \propto a^\dagger aa + (\partial_x a)aa + (\partial_y a^\dagger)aa + a^\dagger(\partial_x a)a + a^\dagger(\partial_y a)a + h.c.$$
$$\propto a_e a_b^\dagger a_d^\dagger + a_e a_b a_s^\dagger + h.c., \quad (9)$$

Here, $a_e$ represents the magnon excitation mode with frequency $\omega_e$, while $a_b$ denotes the skyrmion breathing mode with frequency $\omega_b$. $a_d$ and $a_s$ correspond to the difference-frequency mode with the frequency $\omega_d = \omega_e - \omega_b$ and the sum-frequency mode with frequency $\omega_s = \omega_e + \omega_b$, respectively. This occurrence of sum- and difference- frequency modes during the three-magnon process results in the magnon frequency comb effect.

*C. Manipulation of the magnon frequency comb*

Subsequently, we study the magnon excitation in the system. The dynamics of **m** and **n** can be described by the continuum Landau-Lifshitz-Gilbert (LLG) equation [6,28,29,40],

$$s\dot{\mathbf{m}} + \delta_s \dot{\mathbf{n}} = -(\mathbf{m}\times\mathbf{f}_m + \mathbf{n}\times\mathbf{f}_n) + \alpha s(\mathbf{n}\times\dot{\mathbf{n}}), \quad (10a)$$

$$s\dot{\mathbf{n}} = -(\mathbf{n}\times\mathbf{f}_m) + \alpha\left[\delta_s(\mathbf{n}\times\dot{\mathbf{n}}) + s(\mathbf{n}\times\dot{\mathbf{m}})\right], \quad (10b)$$

where $s = (s_1 + s_2)$ and $\delta_s = (s_1 - s_2)$ with the magnitude of the spin density $s_i = M_i/\gamma_i$, $M_i$ is the magnetization and $\gamma_i = g_i\mu_B/\hbar$ is the gyromagnetic ratio of sublattice $i$. Simply, the damping constants of two layers are considered to be the same. $\mathbf{f}_n = -\delta H/\delta\mathbf{n}$ and $\mathbf{f}_m = -\delta H/\delta\mathbf{m}$ denote the effective fields of **n** and **m**, respectively. After safe and necessary simplifications of insignificant terms, we obtain

$$\mathbf{m} = \frac{sc}{4\sigma}\dot{\mathbf{n}}\times\mathbf{n}, \quad \dot{\mathbf{m}} = \frac{sc}{4\sigma}\ddot{\mathbf{n}}\times\mathbf{n}. \quad (11)$$

It is revealed that the dynamics of **m** is mainly determined by the inter-layer exchange coupling.

We derive the equation of motion for **n** by inserting Eq. (11) into [Eq. (10a)], and obtain

$$\rho \ddot{\mathbf{n}} \times \mathbf{n} + \delta_s \dot{\mathbf{n}} = -\mathbf{n} \times \mathbf{f}_n + \alpha s (\mathbf{n} \times \dot{\mathbf{n}}), \tag{12}$$

where $\rho = s^2/a$ is the constant of inertia with the homogeneous exchange constant $a = \sigma/c$.

By defining a complex field as $\psi_\pm = n_x \pm i n_y$ for the right- (+ sign) and left-handed (− sign) magnons, and linearizing the above equation for $n_x$ and $n_y$, the dispersion and magnon frequency gap $f_{\text{gap}}$ are obtained,

$$f_\pm = \frac{\pm \delta_s + \sqrt{\delta_s^2 + 4\rho[4Ak^2 + 4K']}}{2\rho}, \tag{13a}$$

$$f_{\text{gap}}^\pm = \frac{\pm \delta_s + \sqrt{\delta_s^2 + 16\rho K'}}{2\rho}, \tag{13b}$$

where $k$ is the magnon wave vector, and +/− corresponds to the right-/left-handed magnons. It is demonstrated that $f_{\text{gap}}$ closely depends on $\delta_s$ and the inter-layer exchange coupling $\sigma$. For non-zero $\delta_s$, the left- and right-handed magnons demonstrate different dispersion behaviors and resonance frequencies. Take the left-handed magnons as an example, two $\delta_s$ terms in numerator are with opposite signs, leading to the insensitiveness of the frequency gap to $\delta_s$. Furthermore, besides tuning $\delta_s$, one may adjust the spacer thickness to control the inter-layer coupling and to modulate the frequency gap.

Undoubtedly, the theoretical analysis presented above should be checked by numerical simulations to ensure its reliability, as will be reported in the next section.

### III. MICROMAGNETIC SIMULATED RESULTS and DISCUSSION

In this section, we perform the micromagnetic simulations to check the validity of above theoretical analysis and to further reveal the modulation of the magnon frequency comb in ferrimagnets. Here, the simulations are performed on a ferrimagnetic bilayer GdCo/Co with the size of 500 nm × 500 nm × 2 nm and the cell size of 1 nm × 1 nm × 1 nm using MUMAX3 [44], noting that GdCo and Co are widely used materials.

We set the thickness of each ferromagnetic layer 1nm, the time step of $5 \times 10^{-15}$ s, the intra-sublattice exchange stiffness $A = 10$ pJ/m, the perpendicular magnetic anisotropy constant $K =$ 0.2 MJ/m$^3$, DM coefficient $D = 0.9$ mJ/m$^2$, the gyromagnetic ratio $\gamma_i = g_i\mu_B/\hbar$ with the g-factors $g_1 = 2.2$ and $g_2 = 2$, the damping constants $\alpha_1 = \alpha_2 = 0.001$, the bilinear surface exchange coefficient $\sigma$ ranges from $-10$ mJ/m$^2$ to $-30$ mJ/m$^2$, which are reasonable values for the realistic system. The magnetizations of nine various cases are shown in Table 1, which correspond to nine different $\delta_s$ and $M_d$. Here, the value of $M_d$ keeps decreasing as $\delta_s$ increases, consistent with the experiments in which elevating temperature to modulate $\delta_s$ reduces both the magnetizations of two sub-lattices [31,45]. The magnons are excited by applying a microwave driving field **h**(t) = ($h\sin(2\pi ft)$, 0, 0) on a nanobelt, and they propagate along the x-direction and are scattered by the skyrmion, as depicted in Fig. 1(c).

*A. Effect of $\delta_s$ on the magnon frequency comb*

In ferrimagnets, the demagnetization field and $\delta_s$ can be modulated through tuning the magnetizations $M_1$ and $M_2$, which affect $M_d$ and the skyrmion size. Fig. 2(a) gives the Eq. (7)-calculated and LLG simulated skyrmion radius $R$ for various $\delta_s$, which demonstrates a monotonous decrease of $R$ with the increase of $\delta_s$. It is noted that the skyrmion size is mainly determined by the competition between the magnetic anisotropy term and the DM interaction. Thus, the corrective anisotropy $K'$ is enhanced as $\delta_s$ increases due to the decrease of $M_d$, while the DM interaction hardly be affected, resulting in the gradual decrease of the skyrmion size [31,45].

Subsequently, the effect of $\delta_s$ on the skyrmion breathing mode is investigated through analyzing the magnetic excitation spectrum. Here, a sinc-function field **h**$_0$(t) = ($h_0\text{sinc}(2\pi f_0 t)$, 0, 0) with the amplitude $h_0 = 50$ mT and frequency $f_0 = 900$ GHz is applied over the whole film to induce the system response. Then, the internal spectrum is obtained by performing a fast Fourier transform (FFT) for each Néel vector **n**, and the corresponding results are shown in Fig. 2(b). Specifically, the spectra of the in-plane component $n_x$ or $n_y$ depict the ferrimagnetic resonance frequency, representing the magnon frequency gap, whereas the spectrum of the out-of-plane component $n_z$ corresponds to the skyrmion breathing mode. The primary peak of the breathing mode gradually shifts towards the high frequency side as $\delta_s$ increases, demonstrating the effective modulation of the breathing mode through tuning $\delta_s$. Generally, it is expected that a particle with small mass has a pronounced response to external fluctuations [41]. As a result, the breathing mode frequency increases when the radius and effective mass of the skyrmion are decreased, as summarized in Fig. 2(a). Importantly, the frequency alternation reaches up to ~50 GHz, providing an opportunity to elaborately manipulate the magnon frequency comb teeth.

It has been established that generating magnon frequency comb requires a substantial driving force [12], and a microwave field $\mathbf{h}(t) = (h\sin(2\pi ft), 0, 0)$ with a large amplitude $h =$ 100 mT and frequency $f =$ 500 GHz is applied to excite magnons. To investigate the nonlinear magnon processes, we analyze the time-dependent Néel vector $\mathbf{n}$ around the skyrmion region using FFT. Fig. 3(a) illustrates the response distribution in the frequency space for various $\delta_s$. Within the range of $\delta_s$ from $-1.24 \times 10^{-7}$ Js/m$^3$ to $+1.24 \times 10^{-7}$ Js/m$^3$, a series of peaks arise and take shape as a comb. The spacing between the peaks increases with the increasing $\delta_s$, while the number of magnon coherent modes decreases. This method of tuning the coherent frequency spacing is feasible since $\delta_s$ can be stably adjusted. Furthermore, the magnitude of the frequency peak depends on the discrepancy between the driving frequency and the ferrimagnetic resonance frequency. For example, as $\delta_s$ increases, the magnon energy decreases and more magnons are excited during the magnon nonlinear process. Thus, the magnitudes of the frequency peaks in combs are generally enhanced as the ferrimagnetic resonance frequency (dashed lines) approaches to the driving frequency, as shown in Fig. 3(a).

The magnon dispersion are simulated using FFT for $\mathbf{n}$ in the magnon transmission region to estimate the magnon frequency gap. Fig. 4 presents the simulated dispersion of the left- and right-handed magnons for various $\delta_s$ for $\sigma = -10$ mJ/m$^2$, which coincides well with the theoretical prediction in Eq. (13). For compensated ferrimagnet with zero $\delta_s$, the dispersions of the left- and right-handed magnons overlap with each other with a frequency gap around ~255 GHz [29], as shown in Fig. 4(b). The overlap of the dispersions is broken for non-zero $\delta_s$. For a positive $\delta_s$, the frequency gap of the right-handed magnons decreases, while that of the left-handed magnons is enhanced, as shown in Fig. 4(a). One notes that the right-handed magnons are with an energy generally lower than the left-handed magnons for a positive $\delta_s$, resulting in a smaller frequency gap. The same mechanism also works for negative $\delta_s$, as demonstrated in Fig. 4(c) which shows that the left-handed magnons are with a frequency gap lower that the right-handed magnons.

The Eq. (13)-calculated and LLG simulated frequency gaps $f_{gap}$ for various $\delta_s$ are summarized in Fig. 3(b). The analytical results are well consistent with the simulations, while the quantitative deviation for large $|\delta_s|$ may attribute to the ignorance of some related higher-order terms of **m** in theory. It is clearly shown that $f_{gap}$ of the left-handed magnon gradually decreases with the increase of $\delta_s$, while that of the right-handed magnons increases rather quickly, resulting in the $f_{gap}$ discrepancy between the left- and right-handed magnons for non-zero $\delta_s$.

As a matter of fact, the $\delta_s$-induced $f_{gap}$ discrepancy affects the frequency domain of the comb because the lowest coherent frequency is limited by $f_{gap}$. In other words, the magnons can be excited and transferred only when its frequency is larger than $f_{gap}$, while only the skyrmion breathing mode $f_b$ exists below the frequency gap. To investigate the effect of $\delta_s$ on the coherence modes, we calculated the spectra of magnons and present the results in Fig. 5. Taking $\delta_s = -1.24 \times 10^{-7}$ Js/m$^3$ as an example shown in Fig. 5(a), the lowest coherent frequency of the left-handed magnons is ~276 GHz, while that of the right-handed magnons is ~190 GHz. Thus, $\delta_s$ impacts not only the frequency comb teeth but also the coherence modes of the right- and left-handed magnons, as also shown in Fig. 5(b).

Furthermore, it is worth noting that there is a threshold value of the exciting field to generate the frequency comb. The simulated response frequency ($f_r$) of the system as a function of the driving amplitude for various $\delta_s$ are presented in Fig. 6. The frequency comb can be obviously observed when the exciting field is larger than 41.1 mT for $\delta_s = -1.24 \times 10^{-7}$ Js/m$^3$. Furthermore, the threshold value significantly decreases with the increase of $\delta_s$, which attributes to the fact that the resonant frequency of ferrimagnets gradually approaches to the driving frequency. Thus, the excitation of magnons is enhanced as $\delta_s$ increases, and the frequency comb could be generated even under a weak exciting field.

## B. Modulation of the frequency gap

The $\delta_s$-dependent skyrmion size and frequency gap of magnons have been clarified, allowing one to modulate the teeth and coherent mode of the comb through tuning $\delta_s$. Moreover, the frequency gap also depends on the inter-layer coupling, as predicted in Eq. (13), which deserves to be further revealed considering the profound significance of extending the lowest coherent mode of the comb for future applications and the controllability of the inter-layer coupling through tuning the spacer depth.

Here, the effect of the inter-layer coupling on $f_{\text{gap}}$ is investigated through analyzing the surface exchange coefficient $\sigma$-dependent magnon dispersion. For simplicity, a sinc-function field with a frequency ~900GHz is applied to excite the magnon modes in compensated ferrimagnet $\delta_s = 0$. Fig. 7 presents the simulated and calculated magnon dispersion for various $\sigma$, which exhibits well consistence between the simulations and theory. With the enhancement of the inter-layer coupling, the frequency gap is significantly widened. This phenomenon attributes to the fact that the enhanced inter-layer exchange field suppresses excitation of low energy magnons, similar to the case of antiferromagnets [46]. Thus, one may control the spacer depth and in turn tune the inter-layer coupling to manipulate the lowest coherent mode of the comb. Additionally, both the frequency gaps of the left- and right-handed magnons are enhanced as the inter-layer coupling increases, which could be used to alleviate possible interference of unwanted low-frequency magnon modes.

At last, we investigate the dependence of the magnon frequency comb on the inter-layer coupling and the driving frequency. The simulated response frequency as a function of driving frequency for various $\sigma$ are summarized in Fig. 8. For every frequency, the magnon frequency comb can be realized. With the enhancement of the inter-layer coupling, the skyrmion breathing frequency $f_b$ slightly shifts towards the high frequency side, which enlarges the distance between the comb teeth. Moreover, the resonance frequency also increases and approaches to the driving frequency. Thus, the response of the system is strongly enhanced, as shown in Figs. 8(a)-(c). However, the number of the comb teeth could be reduced, noting that magnons cannot exist below the resonance frequency.

*C. Discussion*

So far, we have elucidated the important role of the net angular momentum $\delta_s$ and the inter-layer coupling in modulating the magnon frequency comb in ferrimagnets, which provides valuable insights for material selection and serves as a guide for future experiments. For example, the coherent modes generated by the comb in ferrimagnets could exist from gigahertz to terahertz frequencies, serving as a bridge between microwave and terahertz waves. More importantly, the controllable comb teeth provide a solution for precise synthesis of high frequency magnon modes.

Similar to the earlier report, $\delta_s$ emerges as a essential control parameter in modulating ferrimagnetic dynamics, offering better control through temperature or material composition adjustments [27–31]. Simultaneously, the inter-layer coupling can be modulated by tuning the spacer thickness [39]. Notably, most parameters chosen in this work are comparable to those in GdCo/Co [20,39,45,47], and the actualization of magnon frequency combs in ferrimagnets awaits validation in forthcoming experiments. Furthermore, the magnon frequency comb effect uncovered here exhibits universality over all chiral magnetic models, and the system can be extended from thin films to bulk materials.

## IV. CONCLUSION

In conclusion, we have studied the magnon frequency comb effect in ferrimagnets using analytical methods and numerical simulations. The coherent modes generated by the comb could exist from gigahertz to terahertz frequencies, serving as a bridge between microwave and terahertz waves. As $\delta_s$ increases, the total magnetization and skyrmion size decreases, which enlarges the skyrmion breathing frequency and the distance between the comb teeth. The close dependence of the skyrmion breathing mode on $\delta_s$ allows one to modulate the teeth of the comb through tuning temperature or material composition. Furthermore, the magnon frequency gap is also related to $\delta_s$ and the inter-layer coupling, providing the opportunity to modulate the lowest coherent frequency of the comb through structural control. Thus, this work demonstrates the controllable magnon frequency comb in ferrimagnets and holds significance for broader application of ferrimagnets and magnon frequency comb.

**Acknowledgment**

The work is supported by the Natural Science Foundation of China (Grants No. U22A20117, No. 52371243, No. 51971096, No. 92163210, and No. 51721001), the Guangdong Basic and Applied Basic Research Foundation (Grant No. 2022A1515011727).

**References:**

[1] S. T. Cundiff and J. Ye, *Colloquium: Femtosecond Optical Frequency Combs*, Rev. Mod. Phys. **75**, 325 (2003).

[2] P. Yan, X. S. Wang, and X. R. Wang, *All-Magnonic Spin-Transfer Torque and Domain Wall Propagation*, Phys. Rev. Lett. **107**, 177207 (2011).

[3] O. Lee, K. Yamamoto, M. Umeda, C. W. Zollitsch, M. Elyasi, T. Kikkawa, E. Saitoh, G. E. W. W. Bauer, and H. Kurebayashi, *Nonlinear Magnon Polaritons*, Phys. Rev. Lett. **130**, 046703 (2023).

[4] R. Mondal and L. Rózsa, *Inertial Spin Waves in Ferromagnets and Antiferromagnets*, Phys. Rev. B **106**, 134422 (2022).

[5] J. Lan, W. Yu, R. Wu, and J. Xiao, *Spin-Wave Diode*, Phys. Rev. X **5**, 041049 (2015).

[6] T. T. Liu et al., *Handedness Filter and Doppler Shift of Spin Waves in Ferrimagnetic Domain Walls*, Phys. Rev. B **105**, 214432 (2022).

[7] H. Wang et al., *Long-Distance Coherent Propagation of High-Velocity Antiferromagnetic Spin Waves*, Phys. Rev. Lett. **130**, 096701 (2023).

[8] X. Liang, Z. Wang, P. Yan, and Y. Zhou, *Nonreciprocal Spin Waves in Ferrimagnetic Domain-Wall Channels*, Phys. Rev. B **106**, 224413 (2022).

[9] V. A. S. V. Bittencourt, C. A. Potts, Y. Huang, J. P. Davis, and S. Viola Kusminskiy, *Magnomechanical Backaction Corrections Due to Coupling to Higher-Order Walker Modes and Kerr Nonlinearities*, Phys. Rev. B **107**, 144411 (2023).

[10] Z. Wang, Y. Cao, and P. Yan, *Goos-Hänchen Effect of Spin Waves at Heterochiral Interfaces*, Phys. Rev. B **100**, 064421 (2019).

[11] Z. R. Yan, Y. W. Xing, and X. F. Han, *Magnonic Skin Effect and Magnon Valve Effect in an Antiferromagnetically Coupled Heterojunction*, Phys. Rev. B **104**, L020413 (2021).

[12] Z. Wang, H. Y. Yuan, Y. Cao, Z. X. Li, R. A. Duine, and P. Yan, *Magnonic Frequency Comb through Nonlinear Magnon-Skyrmion Scattering*, Phys. Rev. Lett. **127**, 037202 (2021).

[13] H. Y. Yuan, Y. Cao, A. Kamra, R. A. Duine, and P. Yan, *Quantum Magnonics: When Magnon Spintronics Meets Quantum Information Science*, Phys. Rep. **965**, 1 (2022).

**References:**

[1] S. T. Cundiff and J. Ye, *Colloquium: Femtosecond Optical Frequency Combs*, Rev. Mod. Phys. **75**, 325 (2003).

[2] P. Yan, X. S. Wang, and X. R. Wang, *All-Magnonic Spin-Transfer Torque and Domain Wall Propagation*, Phys. Rev. Lett. **107**, 177207 (2011).

[3] O. Lee, K. Yamamoto, M. Umeda, C. W. Zollitsch, M. Elyasi, T. Kikkawa, E. Saitoh, G. E. W. W. Bauer, and H. Kurebayashi, *Nonlinear Magnon Polaritons*, Phys. Rev. Lett. **130**, 046703 (2023).

[4] R. Mondal and L. Rózsa, *Inertial Spin Waves in Ferromagnets and Antiferromagnets*, Phys. Rev. B **106**, 134422 (2022).

[5] J. Lan, W. Yu, R. Wu, and J. Xiao, *Spin-Wave Diode*, Phys. Rev. X **5**, 041049 (2015).

[6] T. T. Liu et al., *Handedness Filter and Doppler Shift of Spin Waves in Ferrimagnetic Domain Walls*, Phys. Rev. B **105**, 214432 (2022).

[7] H. Wang et al., *Long-Distance Coherent Propagation of High-Velocity Antiferromagnetic Spin Waves*, Phys. Rev. Lett. **130**, 096701 (2023).

[8] X. Liang, Z. Wang, P. Yan, and Y. Zhou, *Nonreciprocal Spin Waves in Ferrimagnetic Domain-Wall Channels*, Phys. Rev. B **106**, 224413 (2022).

[9] V. A. S. V. Bittencourt, C. A. Potts, Y. Huang, J. P. Davis, and S. Viola Kusminskiy, *Magnomechanical Backaction Corrections Due to Coupling to Higher-Order Walker Modes and Kerr Nonlinearities*, Phys. Rev. B **107**, 144411 (2023).

[10] Z. Wang, Y. Cao, and P. Yan, *Goos-Hänchen Effect of Spin Waves at Heterochiral Interfaces*, Phys. Rev. B **100**, 064421 (2019).

[11] Z. R. Yan, Y. W. Xing, and X. F. Han, *Magnonic Skin Effect and Magnon Valve Effect in an Antiferromagnetically Coupled Heterojunction*, Phys. Rev. B **104**, L020413 (2021).

[12] Z. Wang, H. Y. Yuan, Y. Cao, Z. X. Li, R. A. Duine, and P. Yan, *Magnonic Frequency Comb through Nonlinear Magnon-Skyrmion Scattering*, Phys. Rev. Lett. **127**, 037202 (2021).

[13] H. Y. Yuan, Y. Cao, A. Kamra, R. A. Duine, and P. Yan, *Quantum Magnonics: When Magnon Spintronics Meets Quantum Information Science*, Phys. Rep. **965**, 1 (2022).

[14] S. Zheng, Z. Wang, Y. Wang, F. Sun, Q. He, P. Yan, and H. Y. Yuan, *Tutorial: Nonlinear Magnonics*, J. Appl. Phys. **134**, 151101 (2023).

[15] Z. Wang, H. Y. Yuan, Y. Cao, and P. Yan, *Twisted Magnon Frequency Comb and Penrose Superradiance*, Phys. Rev. Lett. **129**, 107203 (2022).

[16] Z. Jin, X. Yao, Z. Wang, H. Y. Yuan, Z. Zeng, Y. Cao, and P. Yan, *Nonlinear Topological Magnon Spin Hall Effect*, Phys. Rev. Lett. **131**, 166704 (2023).

[17] X. Yao, Z. Jin, Z. Wang, Z. Zeng, and P. Yan, *Terahertz Magnon Frequency Comb*, Phys. Rev. B **108**, 134427 (2023).

[18] N. Nagaosa and Y. Tokura, *Topological Properties and Dynamics of Magnetic Skyrmions*, Nat. Nanotechnol. **8**, 899 (2013).

[19] W. Jiang et al., *Direct Observation of the Skyrmion Hall Effect*, Nat. Phys. **13**, 162 (2017).

[20] S. Woo et al., *Deterministic Creation and Deletion of a Single Magnetic Skyrmion Observed by Direct Time-Resolved X-Ray Microscopy*, Nat. Electron. **1**, 288 (2018).

[21] J. Xia, X. Zhang, X. Liu, Y. Zhou, and M. Ezawa, *Universal Quantum Computation Based on Nanoscale Skyrmion Helicity Qubits in Frustrated Magnets*, Phys. Rev. Lett. **130**, 106701 (2023).

[22] Y. Liu, T. T. Liu, Z. P. Hou, D. Y. Chen, Z. Fan, M. Zeng, X. B. Lu, X. S. Gao, M. H. Qin, and J.-M. Liu, *Dynamics of Hybrid Magnetic Skyrmion Driven by Spin-Orbit Torque in Ferrimagnets*, Appl. Phys. Lett. **122**, 172405 (2023).

[23] Z. Jin, C. Y. Meng, T. T. Liu, D. Y. Chen, Z. Fan, M. Zeng, X. B. Lu, X. S. Gao, M. H. Qin, and J. M. Liu, *Magnon-Driven Skyrmion Dynamics in Antiferromagnets: Effect of Magnon Polarization*, Phys. Rev. B **104**, 054419 (2021).

[24] Y. Liu et al., *Spin-Wave-Driven Skyrmion Dynamics in Ferrimagnets: Effect of Net Angular Momentum*, Phys. Rev. B **106**, 064424 (2022).

[25] J. Iwasaki, A. J. Beekman, and N. Nagaosa, *Theory of Magnon-Skyrmion Scattering in Chiral Magnets*, Phys. Rev. B **89**, 064412 (2014).

[26] C. Schütte and M. Garst, *Magnon-Skyrmion Scattering in Chiral Magnets*, Phys. Rev. B **90**, 094423 (2014).

[27] S. K. Kim, G. S. D. Beach, K.-J. Lee, T. Ono, T. Rasing, and H. Yang, *Ferrimagnetic*

*Spintronics*, Nat. Mater. **21**, 24 (2022).

[28] S. K. Kim, K. J. Lee, and Y. Tserkovnyak, *Self-Focusing Skyrmion Racetracks in Ferrimagnets*, Phys. Rev. B **95**, 140404(R) (2017).

[29] S. K. Kim, K. Nakata, D. Loss, and Y. Tserkovnyak, *Tunable Magnonic Thermal Hall Effect in Skyrmion Crystal Phases of Ferrimagnets*, Phys. Rev. Lett. **122**, 057204 (2019).

[30] Y. Hirata et al., *Vanishing Skyrmion Hall Effect at the Angular Momentum Compensation Temperature of a Ferrimagnet*, Nat. Nanotechnol. **14**, 232 (2019).

[31] K. J. Kim et al., *Fast Domain Wall Motion in the Vicinity of the Angular Momentum Compensation Temperature of Ferrimagnets*, Nat. Mater. **16**, 1187 (2017).

[32] T. T. Liu, Y. F. Hu, Y. Liu, Z. J. Y. Jin, Z. H. Tang, and M. H. Qin, *Domain Wall Dynamics Driven by a Circularly Polarized Magnetic Field in Ferrimagnet: Effect of Dzyaloshinskii–Moriya Interaction*, Rare Met. **41**, 3815 (2022).

[33] C. R. Zhao, Y. X. Wei, T. T. Liu, and M. H. Qin, *Dynamics of Ferrimagnetic Domain Walls Driven by Sinusoidal Microwave Magnetic Field*, Acta Phys. Sin. **72**, 208502 (2023).

[34] V. P. Kravchuk, D. D. Sheka, U. K. Rößler, J. Van Den Brink, and Y. Gaididei, *Spin Eigenmodes of Magnetic Skyrmions and the Problem of the Effective Skyrmion Mass*, Phys. Rev. B **97**, 064403 (2018).

[35] S. Komineas and P. E. Roy, *Breathing Skyrmions in Chiral Antiferromagnets*, Phys. Rev. Res. **4**, 033132 (2022).

[36] B. F. McKeever, D. R. Rodrigues, D. Pinna, A. Abanov, J. Sinova, and K. Everschor-Sitte, *Characterizing Breathing Dynamics of Magnetic Skyrmions and Antiskyrmions within the Hamiltonian Formalism*, Phys. Rev. B **99**, 054430 (2019).

[37] X. S. Wang, H. Y. Yuan, and X. R. Wang, *A Theory on Skyrmion Size*, Commun. Phys. **1**, 31 (2018).

[38] Z. Jin et al., *Dynamics of Antiferromagnetic Skyrmions in the Absence or Presence of Pinning Defects*, Phys. Rev. B **102**, 054419 (2020).

[39] J. Chatterjee, D. Polley, A. Pattabi, H. Jang, S. Salahuddin, and J. Bokor, *RKKY Exchange Bias Mediated Ultrafast All-Optical Switching of a Ferromagnet*, Adv. Funct.

Mater. **32**, 2107490 (2022).

[40] D. H. Kim, S. H. Oh, D. K. Lee, S. K. Kim, and K. J. Lee, *Current-Induced Spin-Wave Doppler Shift and Attenuation in Compensated Ferrimagnets*, Phys. Rev. B **103**, 014433 (2021).

[41] H. B. Braun, *Fluctuations and Instabilities of Ferromagnetic Domain-Wall Pairs in an External Magnetic Field*, Phys. Rev. B **50**, 16485 (1994).

[42] V. P. Kravchuk, O. Gomonay, D. D. Sheka, D. R. Rodrigues, K. Everschor-Sitte, J. Sinova, J. Van Den Brink, and Y. Gaididei, *Spin Eigenexcitations of an Antiferromagnetic Skyrmion*, Phys. Rev. B **99**, 184429 (2019).

[43] T. Holstein and H. Primakoff, *Field Dependence of the Intrinsic Domain Magnetization of a Ferromagnet*, Phys. Rev. **58**, 1098 (1940).

[44] A. Vansteenkiste, J. Leliaert, M. Dvornik, M. Helsen, F. Garcia-Sanchez, and B. Van Waeyenberge, *The Design and Verification of MuMax3*, AIP Adv. **4**, 107133 (2014).

[45] S. H. Oh, S. K. Kim, J. Xiao, and K. J. Lee, *Bidirectional Spin-Wave-Driven Domain Wall Motion in Ferrimagnets*, Phys. Rev. B **100**, 174403 (2019).

[46] L. Qiu and K. Shen, *Tunable Spin-Wave Nonreciprocity in Synthetic Antiferromagnetic Domain Walls*, Phys. Rev. B **105**, 094436 (2022).

[47] S. Woo et al., *Current-Driven Dynamics and Inhibition of the Skyrmion Hall Effect of Ferrimagnetic Skyrmions in GdFeCo Films*, Nat. Commun. **9**, 959 (2018).

Table 1. Parameters used in the micromagnetic simulation.

| Index | 1 | 2 | 3 | 4 | 5 | 6 | 7 | 8 | 9 |
|---|---|---|---|---|---|---|---|---|---|
| $M_1$ (kA/m) | 460 | 455 | 450 | 445 | 440 | 435 | 430 | 425 | 420 |
| $M_2$ (kA/m) | 440 | 430 | 420 | 410 | 400 | 390 | 380 | 370 | 360 |
| $\delta_s$ (×10$^{-7}$ J·s/m$^3$) | −1.24 | −0.93 | −0.62 | −0.31 | 0 | 0.31 | 0.62 | 0.93 | 1.24 |
| $M_d$ (×10$^{11}$ A$^2$/m$^2$) | 2.026 | 1.960 | 1.895 | 1.831 | 1.768 | 1.707 | 1.647 | 1.588 | 1.530 |

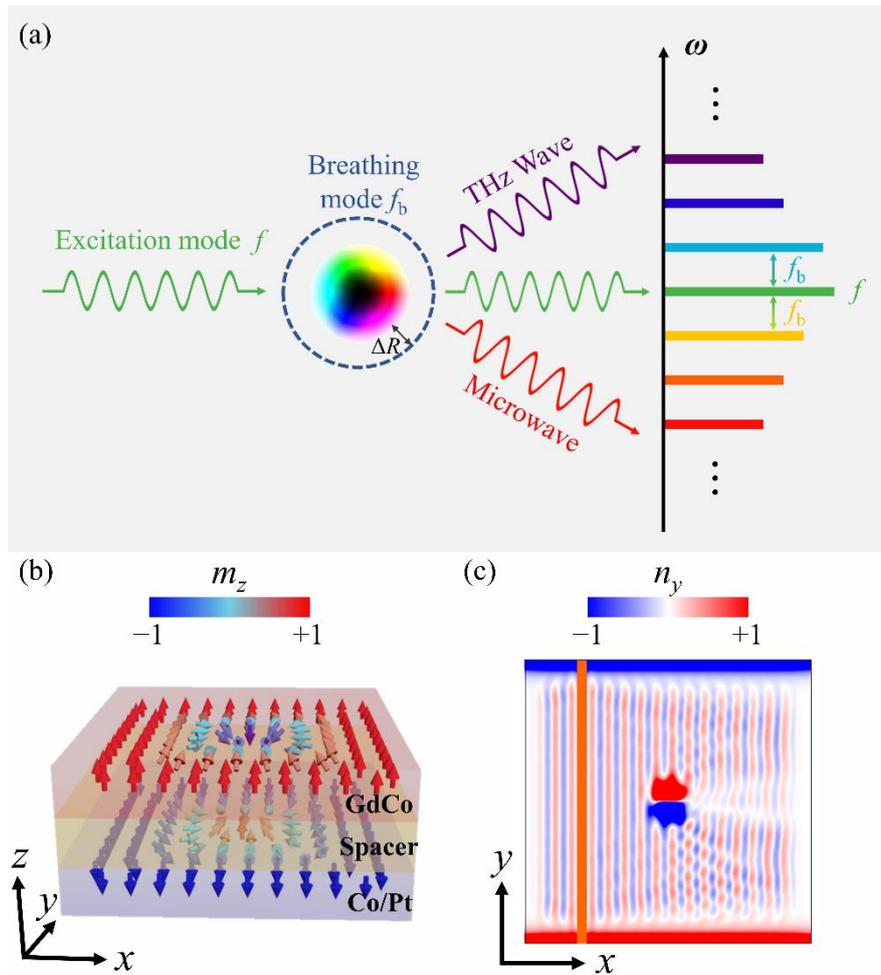

Fig. 1. (a) Schematic illustration of interactions between the ferrimagnetic skyrmion and magnons, and the resulted magnon frequency comb, and (b) sketch of a synthetic ferrimagnetic system with a Néel skyrmion, and (c) snapshot of the interaction between propagating magnons and skyrmion in ferrimagnet.

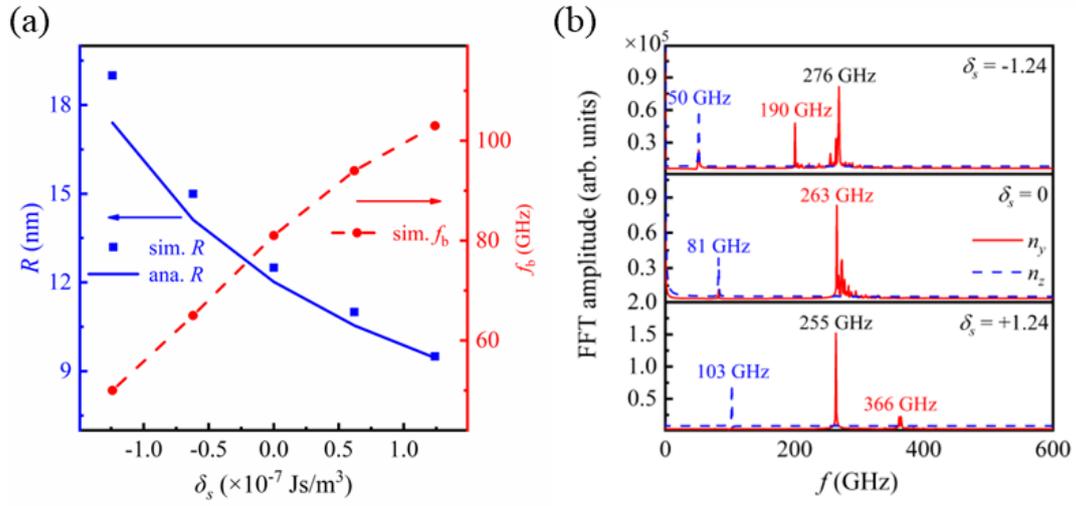

Fig. 2. (a) The analytically calculated (lines) and simulated (symbols) skyrmion radius ($R$) and the skyrmion breathing frequency ($f_b$) as functions of the net angular momentum $\delta_s$, and (b) the skyrmion breathing mode spectrum for $n_y$ and $n_z$ for $\delta_s = -1.24 \times 10^{-7}$ Js/m$^3$ (top), 0 (middle), and $1.24 \times 10^{-7}$ Js/m$^3$ (bottom).

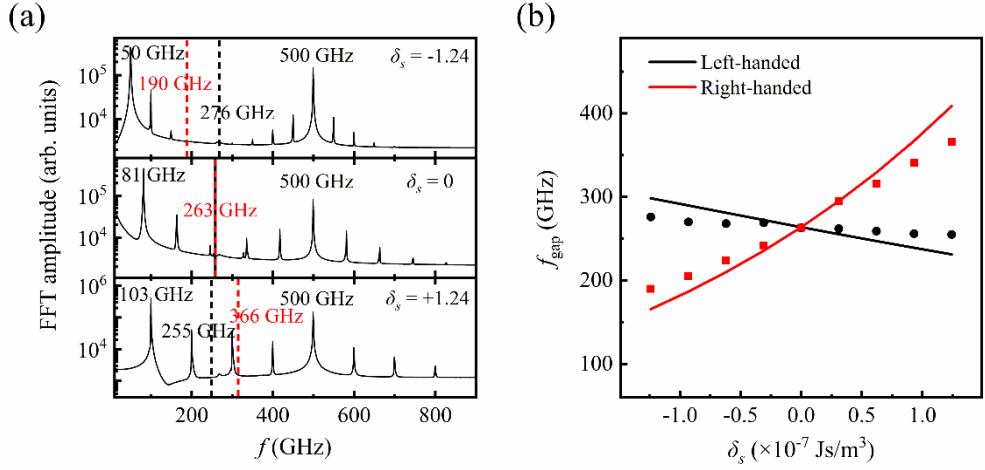

Fig. 3. (a) The magnon spectrum for $\delta_s = -1.24 \times 10^{-7}$ Js/m$^3$ (top), 0 (middle), and $1.24 \times 10^{-7}$ Js/m$^3$ (bottom), respectively. The black (red) dashed line indicates the frequency gap of the left- (right-) handed magnons. (b) The calculated (lines) and simulated (symbols) frequency gap ($f_{\text{gap}}$) as a function of $\delta_s$ with the interlayer coupling $\sigma = -10$ mJ/m$^2$ for different magnon mode handedness.

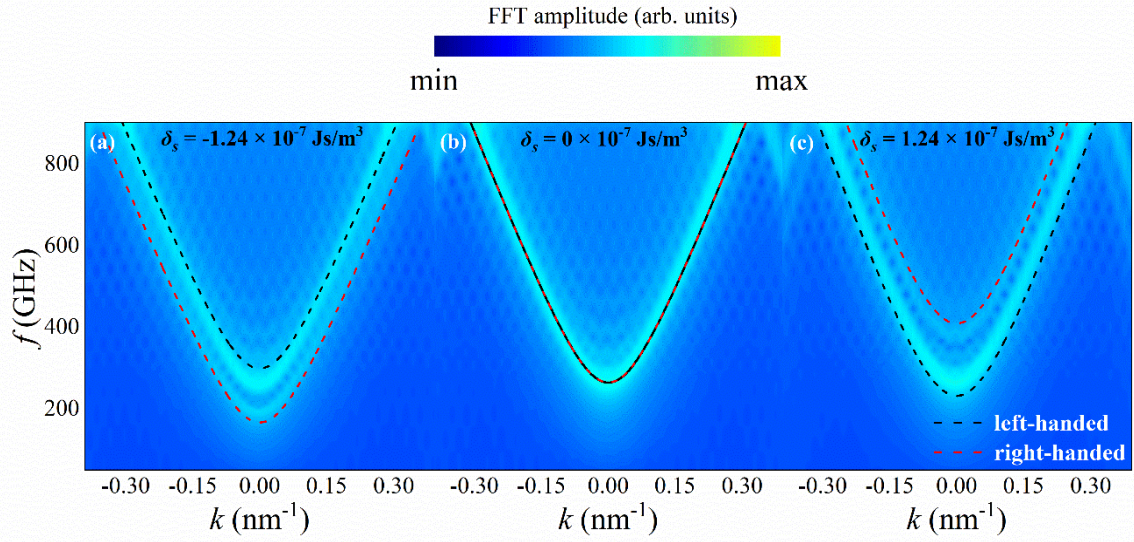

Fig. 4. Ferrimagnetic magnon dispersion for (a) $\delta_s = -1.24 \times 10^{-7}$ Js/m$^3$, (b) $\delta_s = 0$ and (c) $\delta_s = -1.24 \times 10^{-7}$ Js/m$^3$ with the interlayer coupling $\sigma = -10$ mJ/m$^2$. Dash lines show the analytical results.

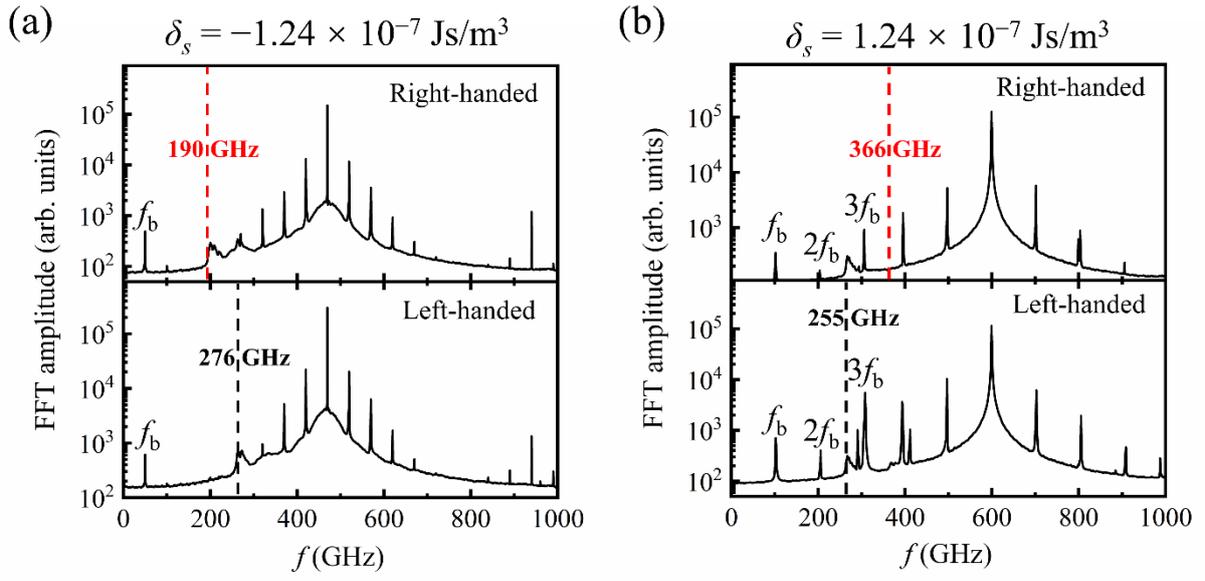

Fig. 5. The magnon spectrum for (a) $\delta_s = -1.24 \times 10^{-7}$ Js/m$^3$ and (b) $1.24 \times 10^{-7}$ Js/m$^3$ with different magnon handednesses. The black (red) dashed line indicates the frequency gap of the left- (right-) handed magnons.

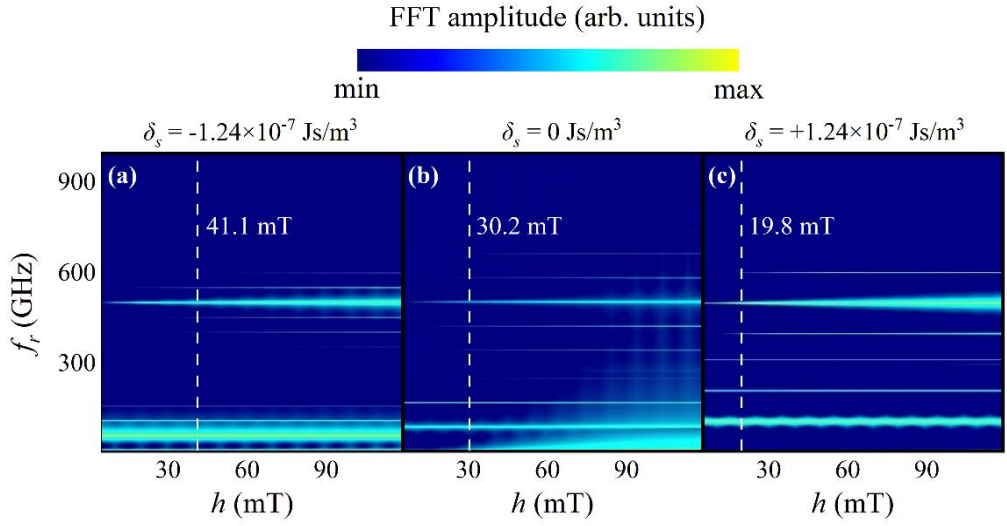

Fig. 6. The response frequency ($f_r$) of the system as a function of the microwave driving amplitude ($h$) for (a) $\delta_s = -1.24 \times 10^{-7}$ Js/m$^3$, (b) $\delta_s = 0$ Js/m$^3$, and (c) $\delta_s = +1.24 \times 10^{-7}$ Js/m$^3$. The driving frequency is fixed at 500 GHz.

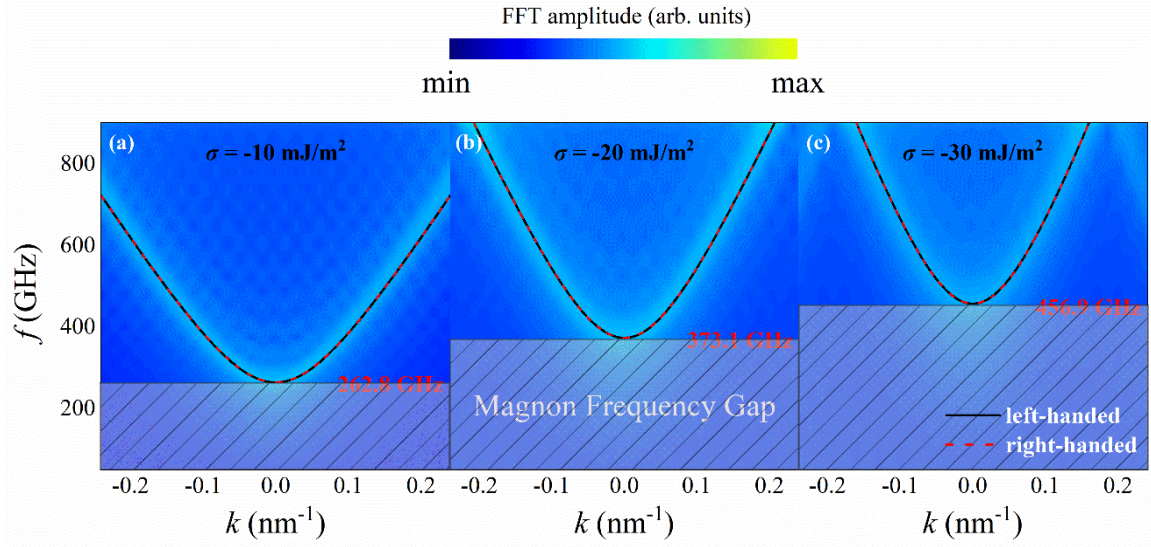

Fig. 7. Ferrimagnetic magnon dispersion for (a) $\sigma = -10$ mJ/m$^2$, (b) $\sigma = -20$ mJ/m$^2$ and (c) $\sigma = -30$ mJ/m$^2$ for $\delta_s = 0$ Js/m$^3$.

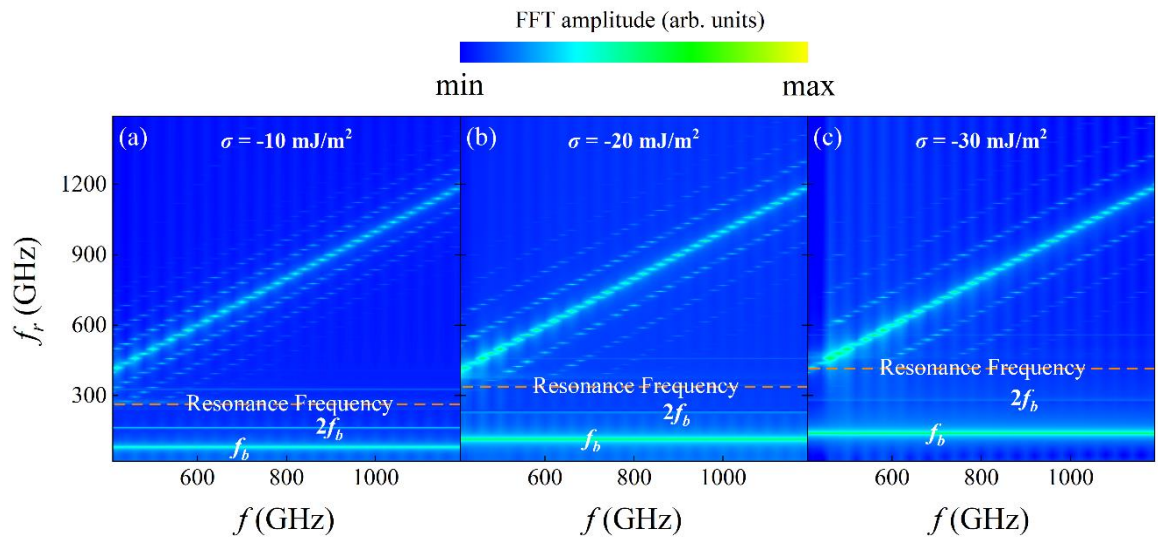

Fig. 8. The response frequency ($f_r$) of the system as a function of the microwave driving frequency ($f$) for (a) $\sigma = -10$ mJ/m$^2$, (b) $\sigma = -20$ mJ/m$^2$, and (c) $\sigma = -30$ mJ/m$^2$.